\documentclass[pra,showpacs,floatfix,superscriptaddress,twocolumn,aps,10]{revtex4-2}
\usepackage{xcolor}
\usepackage{tikz}
\usepackage{polyhedra}
\usepackage{extarrows}
\usepackage{caption}
\usepackage[inline]{enumitem}
\usepackage{amsmath}
\usepackage{amssymb} 
\usepackage{parskip}
\usepackage{comment}
\numberwithin{equation}{section} 
\usepackage[version=4]{mhchem}  
\usepackage{ragged2e}

\usepackage{graphicx}
\usepackage{float}
\usepackage{subcaption}
\usepackage{dcolumn}
\usepackage{bm}
\usepackage[colorlinks=true, allcolors=blue]{hyperref}
\usepackage{cleveref}
\Crefname{figure}{Fig.}{Figs.}

\begin{document}

\preprint{APS/123-QED}

\title{Cumulant dynamics of the ferromagnetic dipolar $XY$ model}

    \author{Jinxin Yang}
	\affiliation{Department of Physics and Arnold Sommerfeld Center for Theoretical Physics (ASC), Ludwig Maximilian University of Munich, 80333 Munich, Germany}

    \author{Alberto R.~Cavallar}
    \affiliation{Department of Physics and Arnold Sommerfeld Center for Theoretical Physics (ASC), Ludwig Maximilian University of Munich, 80333 Munich, Germany}
    \affiliation{Munich Center for Quantum Science and Technology (MCQST), 80799 Munich, Germany}
    
    \author{Lode Pollet}
	\email[]{Lode.Pollet@lmu.de}
	\affiliation{Department of Physics and Arnold Sommerfeld Center for Theoretical Physics (ASC), Ludwig Maximilian University of Munich, 80333 Munich, Germany}
    \affiliation{Munich Center for Quantum Science and Technology (MCQST), 80799 Munich, Germany}
   
\date{\today}

\begin{abstract}
We study the second-order truncated cumulant dynamics for the preparation of dynamical states in the dipolar ferromagnetic $XY$ model on the square lattice for system sizes up to linear size $L=60$. The observations lead to a transparent picture in which a coherent region of finite extent and finite magnetization forms, which is possible even for diabatic but sufficiently slow ramps. These two scales are functions of the ramp constant, the initial entropy of the system, and time. Our findings put strong requirements on future experiments addressing dipolar $XY$ equilibrium physics in bigger system sizes.
\end{abstract}

\maketitle

\section{Introduction}

The Mermin--Wagner theorem forbids the spontaneous breaking of a continuous
symmetry such as $U(1)$ at finite temperature in two dimensions for sufficiently
short-ranged interactions. For couplings decaying as a power law
$1/r^{\alpha}$, the theorem does not apply for $\alpha < 4$~\cite{Defenu2023LongRangeInteracting}.
Addressing the issue of off-diagonal long-range order (ODLRO) with dipolar interactions ($\alpha=3$) has only recently become possible experimentally: Chen \textit{et al.}~\cite{Chen2023ContinuousSymmetryBreaking} hold arrays of Rydberg atoms in optical tweezers~\cite{Bernien2017RydbergSimulator} whose resonant dipolar exchange between two Rydberg levels implements a two-dimensional $XY$ model. Arrays consisting of up to $N=100$ atoms were prepared in a N\'eel-like product state in a strong staggered magnetic field, which was ramped down consequently in a near-adiabatic fashion. Chen \textit{et al.} measured the site-resolved in-plane correlator $C^{xx}\!\left(d\right)$, which was seen to flatten out to the largest accessible separations on a time scale of $t\sim 1-2\,\mu$s. The system was however unable to maintain this plateau at later times, leading to a profile which decays approximately linearly at $t > 5\,\mu$s. The elementary excitations have been characterized separately by quench
spectroscopy~\cite{Chen2025QuenchSpectroscopy}.

The equilibrium properties of the ferromagnetic dipolar $XY$ model are under very good theoretical control. Linear spin wave theory (LSWT) gives a quantitatively accurate description at low enough temeperatures~\cite{Peter2012AnomalousDipolar}. Sbierski \textit{et
al.}~\cite{Sbierski2024Magnetism} used quantum Monte Carlo simulations to obtain the thermodynamic phase diagram and its critical properties. They also simulated the experimental system, and found that a plateau of the experimentally observed magnitude cannot occur in equilibrium at the initial entropy and 
size of the experiment. The linearly decaying curve at later times, by contrast, was found to be compatible with quantum Monte Carlo simulations and its unusual shape is explained by the open boundary conditions that matter strongly for the linear system size, $L=10$.

\begin{figure}[t]
\centering
\includegraphics[width=0.480\textwidth]{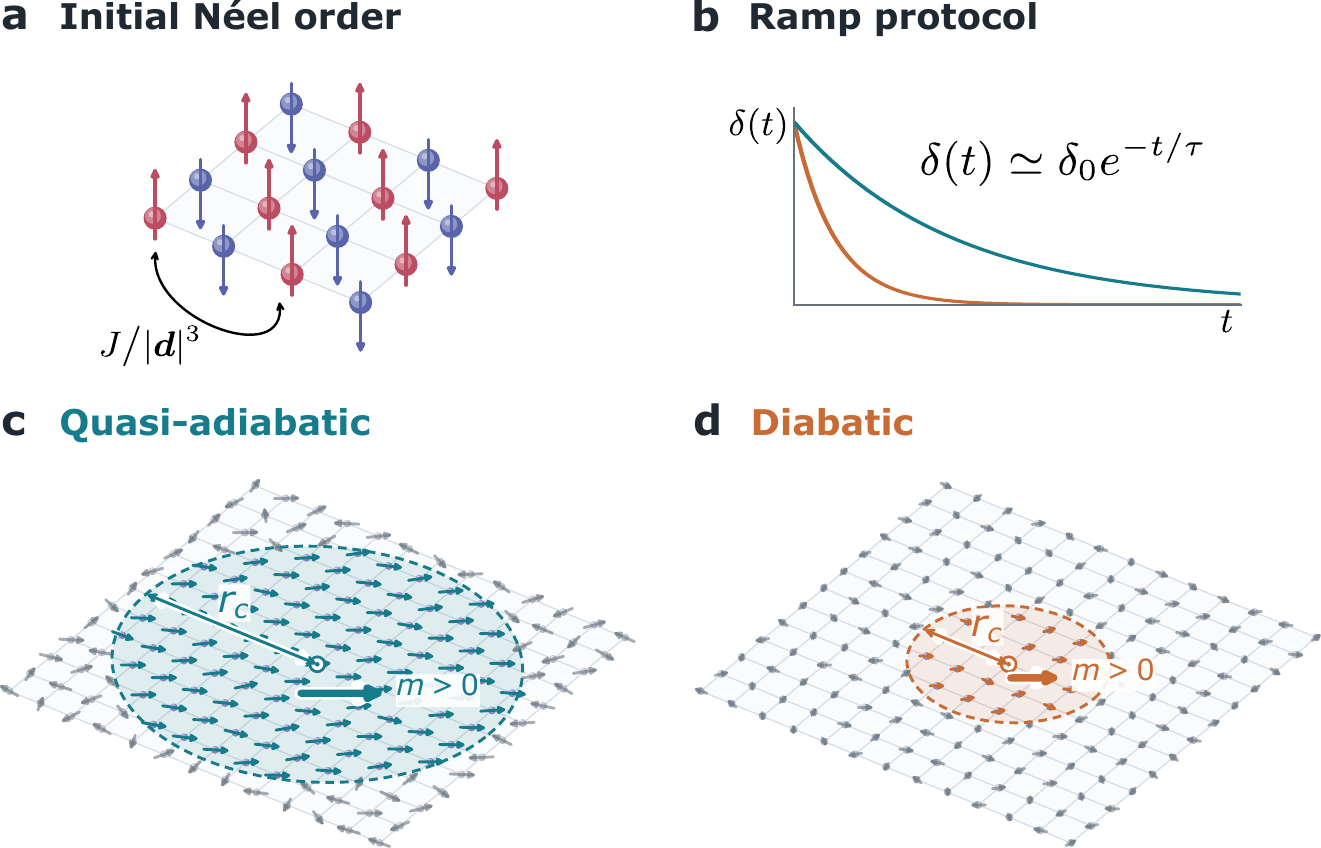}
\caption{\justifying The dynamics of the preparation protocol starting from a N\'eel ordered state under ferromagnetic dipolar $XY$ interactions (a) upon exponentially ramping down the external staggered field (b) leads to a coherent region with finite spontaneous magnetization $m$ over a length scale $r_c$ depending on adiabaticity (c-d).}
\label{fig:sketch}
\end{figure}

In this work we investigate the dynamics of the preparation protocol and go to larger system sizes ($L\sim60$), which are becoming experimentally available. Because it is not possible to obtain exact dynamics for a large quantum system at long times, approximations are inevitable: We use a second-order cumulant truncation for
the spin
dynamics~\cite{Fricke1996CorrelationHierarchy,Akbari2012BBGKY,Pucci2016BBGKY,%
Lackner2015TwoRDM}, which propagates the one- and two-point functions and
discards the connected three-point function. 
This has an $\mathcal{O}\!\left(N^{2}\right)$ memory cost and an $\mathcal{O}\!\left(N^{3}\right)$ operation cost scaling, bringing lattices of size $L\sim 60$ within reach.
We will establish which conservation laws survive the truncation, though the full dynamics at large times is quantitatively questionable.  Qualitative features are however robust, from which we can deduce a transparent model and which predicts that there is a certain length scale over which the system is coherent and has a finite spontaneous magnetization (see Fig.~\ref{fig:sketch}). Outside this region the correlations are paramagnetic and decay therefore in dipolar fashion. The coherence length and the spontaneous magnetization are set by the initial entropy, the ramp constant, and the time. Our findings imply that equilibrium physics can only be studied if the duration of experiment can be extended significantly. 

The paper is organized as follows. Section~\ref{sec:Hamiltonian and Phase Boundary} defines the model
and discusses its phase diagram. 
Section~\ref{sec:Preparation Protocol and Initial State} discusses the initial state preparation.
Section~\ref{sec:Cumulant Dynamics} derives the
second-order cumulant equations.
Section~\ref{sec:Dependence of Plateau} shows the results and interprets their physical meaning.
Section~\ref{sec:Cross-Validation and Positivity}  records the conservation laws and positivity violations of the cumulant dynamics.
Section~\ref{sec:prep_y} addresses the issue of the spectrum obtained through dynamically prepared states, and, finally, we conclude in Sec.~\ref{sec:Conclusions}.


\section{Hamiltonian and Mean-Field Phase diagram}
\label{sec:Hamiltonian and Phase Boundary}
\noindent

We study the two-dimensional dipolar $XY$ Hamiltonian on the square lattice~\cite{Chen2023ContinuousSymmetryBreaking,Sbierski2024Magnetism}
\begin{equation}
\label{eq:Hamiltonian of dipolar XY model}
\hat{H}=-\frac{J}{2}\sum\limits_{\boldsymbol{r}}\sum\limits_{\boldsymbol{r}^{\prime}\neq\boldsymbol{r}}\frac{\hat{S}^{x}_{\boldsymbol{r}}\hat{S}^{x}_{\boldsymbol{r}^{\prime}}+\hat{S}^{y}_{\boldsymbol{r}}\hat{S}^{y}_{\boldsymbol{r}^{\prime}}}{\left|\boldsymbol{r}-\boldsymbol{r}^{\prime}\right|^{3}}+\delta\sum\limits_{\boldsymbol{r}}\eta_{\boldsymbol{r}}\hat{S}^{z}_{\boldsymbol{r}},
\end{equation}
where $J > 0$ is the ferromagnetic coupling, $\mathbf{r}$ and $\mathbf{r'}$ are lattice coordinates, $\hat{S}$ are spin-1/2 operators, $\delta$ is the strength of an external staggered field, and $\eta_{\boldsymbol{r}}= \pm 1$ on the two sublattices, respectively.
To the best of our knowledge, the mean-field phase diagram has not been derived explicitly in the literature before. It can be found by a straightforward extension of the linear spin-wave theory (LSWT)~\cite{AltlandSimons2023CondensedMatterFieldTheory}, which was studied for this model before~\cite{Peter2012AnomalousDipolar,Chen2023ContinuousSymmetryBreaking,Sbierski2024Magnetism,Chen2025QuenchSpectroscopy}.
The Hamiltonian is invariant under a global $U(1)$ rotation.
The (spontaneous) magnetization can therefore be parameterized as 
\begin{eqnarray}
    \boldsymbol{m}_{\boldsymbol{r}} & = & m\left(\sin{\theta} \cos{\phi} \boldsymbol\, \boldsymbol{e}_{x} + \sin{\theta} \sin{\phi} \,\boldsymbol{e}_{y} -\eta_{\boldsymbol{r}}\cos{\theta} \, \boldsymbol{e}_{z}\right), \nonumber \\
    {} & {} & 0\leq m\leq\tfrac{1}{2}, ~~ 0\leq\theta\leq\tfrac{\pi}{2}, ~~ 0\leq\phi< 2\pi.
\end{eqnarray}
Here, $m$ is the strength of the spontaneous magnetization and $\theta$  the characteristic angle away from the staggered $z$ axis. 
In equilibrium, the phase $\phi$ can always be gauged away.
An easy-plane ferromagnet therefore requires a nonzero value of $m$ and $\theta$, otherwise the phase is a paramagnet.
The dipolar contribution over all lattice sites,~$\gamma$, admits a factorization in terms of the Riemann Zeta and Dirichlet Beta functions~\cite{Peter2012AnomalousDipolar,TravenecSamaj2022Epstein}, yielding
\begin{equation}
\label{eq:definition of gamma in dipolar XY model}
\gamma=\sum\limits_{\boldsymbol{d}\neq\boldsymbol{0}}\frac{1}{\left|\boldsymbol{d}\right|^{3}}= 4 \zeta(\tfrac{3}{2})\beta({\tfrac{3}{2}}) \simeq9.034.
\end{equation}

The energy density of a product-state-Ansatz for the magnetization ($\phi=0$) is
\begin{equation}
e=-\frac{J\gamma}{2}m^{2}\sin^{2}{\theta}-\delta m\cos{\theta}.
\end{equation}
Adding the single-site entropy $-\sum\limits_{\sigma=\pm}p_{\sigma}\ln p_{\sigma}$ with $p_{\pm}=\frac{1}{2}\pm m$ leads to the mean-field free energy density at finite temperature $T$ 
\begin{eqnarray}
\label{eq:free energy density in dipolar XY model}
f & = & -\frac{J\gamma}{2}m^{2}\sin^{2}{\theta}-\delta m\cos{\theta} +T\left[\left(\tfrac{1}{2}+m\right)\ln{\left(\tfrac{1}{2}+m \right)} \right. \nonumber \\
{} & {} & \left. +\left(\tfrac{1}{2}-m\right)\ln{\left(\tfrac{1}{2}-m\right)}\right].
\end{eqnarray}
Its stationary conditions are
\begin{eqnarray}
\label{eq:minimum condition of free energy density}
-J\gamma m\sin^2\theta-\delta\cos\theta+T\ln\frac{1+2m}{1-2m} & =& 0, \nonumber \\ 
m\sin\theta\left(\delta-J\gamma m\cos\theta\right) & = & 0.
\end{eqnarray}
The second equation separates the two phases. In the paramagnet,
\begin{equation}
\label{eq:magnetization of paramagnetism in dipolar XY model}
\text{PM:}\quad\theta=0,\quad m=\frac12\tanh\frac{\delta}{2T},
\end{equation}
and in the ferromagnet,
\begin{equation}
\label{eq:magnetization of ferromagnetism in dipolar XY model}
\text{FM:}\quad\cos{\theta}=\frac{\delta}{J\gamma m},\quad\frac{J\gamma}{T}=\frac{1}{m}\ln{\frac{1+2m}{1-2m}}.
\end{equation}

\begin{figure}[!htbp]
\centering
\includegraphics[width=0.450\textwidth]{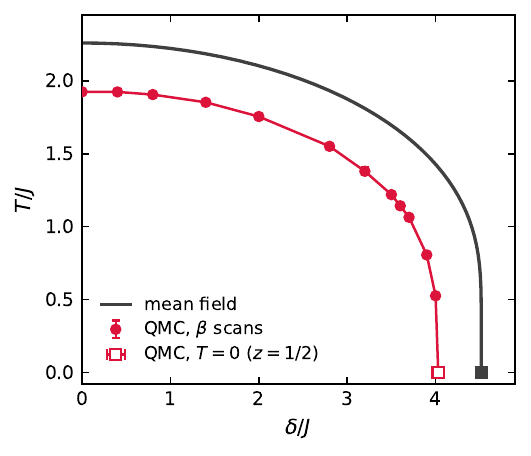}
\caption{\justifying Mean-field equilibrium phase diagram of the two-dimensional dipolar $XY$ model compared with exact quantum Monte Carlo simulations~\cite{Sbierski2024Magnetism}. Monte Carlo error bars are shown but smaller than symbol size. The $XY$-ferromagnetic phase is found in the lower left corner, the paramagnetic phase elsewhere.}
\label{fig:dipolar xy phase diagram numerical result}
\end{figure}

One can find the phase boundary by approaching $\theta\rightarrow0$ in the ferromagnetic phase, with solution 
\begin{equation}
\label{eq:phase boundary equation in dipolar XY model}
\tanh{\frac{\delta}{2T}}=\frac{2\delta}{J\gamma},\quad T_{c}\left(\delta=0\right)=\frac{J\gamma}{4},\quad\delta_{c}\left(T=0\right)=\frac{J\gamma}{2}.
\end{equation}
As shown in Fig.~\ref{fig:dipolar xy phase diagram numerical result} the mean-field phase diagram overestimates the one obtained in quantum Monte Carlo simulations~\cite{Sbierski2024Magnetism} by $\sim 10\%$.

Spin correlation functions $C^{a,b}_{\mathbf{r}, \mathbf{r'}}$ for $a,b \in \{x,y,z \}$ can also be derived straightforwardly.
It was established in Ref.~\cite{Sbierski2024Magnetism} that LSWT remains a good approximation up to $T/J \lesssim 1$ at $\delta = 0$.

For later usage, we also note that a two-body density matrix is reconstructed from the Pauli-product basis~\cite{NielsenChuang2010QuantumInformation} as
\begin{eqnarray}
\rho_{\boldsymbol{r},\boldsymbol{r}^{\prime}} &= & \frac{1}{4}\hat{I}_{2,\boldsymbol{r}}\otimes\hat{I}_{2,\boldsymbol{r}^{\prime}}+\sum\limits_{a}m^{a}_{\boldsymbol{r}}\hat{S}^{a}_{\boldsymbol{r}}\otimes\hat{I}_{\boldsymbol{r}^{\prime}}+\sum\limits_{b}m^{b}_{\boldsymbol{r}^{\prime}}\hat{I}_{\boldsymbol{r}}\otimes\hat{S}^{b}_{\boldsymbol{r}^{\prime}} \nonumber \\ 
{} & {} & +4\sum\limits_{a,b}\left(C^{a,b}_{\boldsymbol{r},\boldsymbol{r}^{\prime}}+m^{a}_{\boldsymbol{r}}m^{b}_{\boldsymbol{r}^{\prime}}\right)\hat{S}^{a}_{\boldsymbol{r}}\otimes\hat{S}^{b}_{\boldsymbol{r}^{\prime}},    
\label{eq:definition of two-body density function of ramped dipolar XY model}
\end{eqnarray}
with $a,b\in\left\lbrace x,y,z\right\rbrace$. 


\section{Initial State preparation and ramp protocol}
\label{sec:Preparation Protocol and Initial State}
\noindent

In Ref.~\cite{Chen2023ContinuousSymmetryBreaking} the system was initially prepared in a system with $\delta_0 \gg J$, followed by an exponentially decaying ramp of the staggered field,
\begin{equation}
\label{eq:external field of ramped dipolar XY model}
\delta\left(t\right)=\begin{cases}
        \delta_{0},&t<0 \\
        \delta_{0}e^{-\frac{t}{\tau}},&t\geq0.
\end{cases}
\end{equation}
The initial system can be modeled as a product state, where on every site we have 
\begin{equation}
\label{eq:initial magnetization of ramped dipolar XY model}
\begin{cases}
    m^{x}_{\boldsymbol{r}}\left(0\right)=m^{y}_{\boldsymbol{r}}\left(0\right)=0 \\
    m^{z}_{\boldsymbol{r}}\left(0\right)=-m\eta_{\boldsymbol{r}}
\end{cases}, \, \, \, \rho_{\boldsymbol{r}}=\frac{1}{2}\hat{I}_{2}+2\boldsymbol{m}_{\boldsymbol{r}}\cdot\boldsymbol{\hat{S}}_{\boldsymbol{r}}.
\end{equation}
The density matrix has two eigenvalues $\frac{1}{2}\pm m\eta_{\boldsymbol{r}}$. The limit $m=1/2$ corresponds to the pure N\'eel state, whereas $m<1/2$ represents a mixed state preparation. The initial strength of the magnetization is determined by the entropy density
\begin{equation}
\label{eq:relation between the entropy density and the magnetization in ramped dipolar XY model}
s=-\left(\tfrac{1}{2}-m\right)\ln{\left(\tfrac{1}{2}-m\right)}-\left(\tfrac{1}{2}+m\right)\ln{\left(\tfrac{1}{2}+m\right)},
\end{equation}
which gives $m\simeq0.411$ for the experimentally inferred value $s=0.30$ used in the comparison below~\cite{Sbierski2024Magnetism}.

Since the initial density matrix corresponds to a product state, all off-site connected correlations are zero, $C^{a,b}({\boldsymbol{r} \neq \boldsymbol{r}^{\prime}})=0$.
However, this does not imply zero on-site fluctuations, as dictated by the spin-$1/2$ algebra, 
$C^{a,b} ({\boldsymbol{r} = \boldsymbol{r}'})=\frac{1}{4}\delta_{a,b}+\frac{i}{2}\sum\limits_{c}\varepsilon_{a,b,c}m^{c}_{\boldsymbol{r}}-m^{a}_{\boldsymbol{r}}m^{b}_{\boldsymbol{r}}
$.
Therefore, $C^{x,x}\left(0\right)=C^{y,y}\left(0\right)=\frac{1}{4}$ and $C^{z,z}\left(0\right)=\frac{1}{4}-m^{2}$.


\section{Cumulant Dynamics: formalism}
\label{sec:Cumulant Dynamics}
\noindent

We perform dynamics under the standard time-dependent many-body equation-of-motion hierarchy~\cite{StefanucciVanLeeuwen2025Nonequilibrium}. 
It is closed at second order by truncating the connected three-point correlation as\,$\big\langle\hat{S}^{a_{1}}_{\boldsymbol{r}_{1}}\hat{S}^{a_{2}}_{\boldsymbol{r}_{2}}\hat{S}^{a_{3}}_{\boldsymbol{r}_{3}}\big\rangle_c=0$, and leads to the cumulant approximation~\cite{Akbari2012BBGKY,Pucci2016BBGKY}
\begin{eqnarray}
\label{eq:fundamental approximation of ramped dipolar XY model}
\big\langle\hat{A}\hat{B}\hat{C}\big\rangle & \simeq & \big\langle\hat{A}\big\rangle\big\langle\hat{B}\hat{C}\big\rangle+\big\langle\hat{B}\big\rangle\big\langle\hat{A}\hat{C}\big\rangle+\big\langle\hat{C}\big\rangle\big\langle\hat{A}\hat{B}\big\rangle \nonumber \\ 
{} & {} & -2\big\langle\hat{A}\big\rangle\big\langle\hat{B}\big\rangle\big\langle\hat{C}\big\rangle.
\end{eqnarray}

The time evolution of the magnetization can be found by taking the expectation value of the Heisenberg equation of motion~\cite{Shimizu2018OutOfEquilibrium}
\begin{eqnarray}
\label{eq:time evolution of magnetization of ramped dipolar XY model}
\frac{\text{d}m^{a}_{\boldsymbol{r}}}{\text{d}t}& = & \frac{J}{2}\sum\limits_{\boldsymbol{s}\neq\boldsymbol{r}}\sum\limits_{b}\frac{1}{\left|\boldsymbol{s}-\boldsymbol{r}\right|^{3}}\left[\varepsilon_{x,a,b}\left(C^{b,x}_{\boldsymbol{r},\boldsymbol{s}}+C^{x,b}_{\boldsymbol{s},\boldsymbol{r}}+2m^{b}_{\boldsymbol{r}}m^{x}_{\boldsymbol{s}}\right) \right. \nonumber \\
{} & {} & + \left. \varepsilon_{y,a,b}\left(C^{b,y}_{\boldsymbol{r},\boldsymbol{s}}+C^{y,b}_{\boldsymbol{s},\boldsymbol{r}}+2m^{b}_{\boldsymbol{r}}m^{y}_{\boldsymbol{s}}\right)\right] \nonumber \\ 
{} & {} & -\delta_{0}e^{-\frac{t}{\tau}}\eta_{\boldsymbol{r}}\sum\limits_{b}\varepsilon_{z,a,b}m^{b}_{\boldsymbol{r}}.
\end{eqnarray}

The connected correlation can be found from the relation
\begin{equation}
\label{eq:time evolution of correlation of ramped dipolar XY model}
\frac{\text{d}C^{a,b}_{\boldsymbol{r},\boldsymbol{r}^{\prime}}}{\text{d}t}=\left\langle\frac{\text{d}\hat{S}^{a}_{\boldsymbol{r}}}{\text{d}t}\hat{S}^{b}_{\boldsymbol{r}^{\prime}}\right\rangle+\left\langle\hat{S}^{a}_{\boldsymbol{r}}\frac{\text{d}\hat{S}^{b}_{\boldsymbol{r}^{\prime}}}{\text{d}t}\right\rangle-\frac{\text{d}m^{a}_{\boldsymbol{r}}}{\text{d}t}m^{b}_{\boldsymbol{r}^{\prime}}-m^{a}_{\boldsymbol{r}}\frac{\text{d}m^{b}_{\boldsymbol{r}^{\prime}}}{\text{d}t}.
\end{equation}
After applying Eq.~\eqref{eq:fundamental approximation of ramped dipolar XY model}, the relation above is closed in terms of $m^{a}_{\boldsymbol{r}}$ and $C^{a,b}_{\boldsymbol{r}_{1},\boldsymbol{r}_{2}}$. 

The cumulant approximation to the equations of motion respects hermiticity. We will also make use of the following symmetries: First, based on the global $O(2)$ symmetry, one can show that 
\begin{equation}
\label{eq:third simplification based on initial condition of ramped dipolar XY model}
C^{x,x}_{\boldsymbol{r},\boldsymbol{r}^{\prime}}=C^{y,y}_{\boldsymbol{r},\boldsymbol{r}^{\prime}} \quad ,\qquad C^{x,y}_{\boldsymbol{r},\boldsymbol{r}^{\prime}}=-C^{x,y}_{\boldsymbol{r}^{\prime},\boldsymbol{r}}\quad ,
\end{equation}
valid at all times.
Second, as a consequence of the N\'eel-like initial condition, $m^{z}_{\boldsymbol{r}}(0)=-m_{0}\,\eta_{\boldsymbol{r}}$ with
$\eta_{\boldsymbol{r}}=e^{i\boldsymbol{Q}\cdot\boldsymbol{r}}$,
$\boldsymbol{Q}=\left(\pi,\pi\right)$, the following quantities are initially zero and remain identically zero throughout the evolution:
\begin{equation}
\label{eq:vanishing components based on initial condition of ramped dipolar XY model}m^{x}_{\boldsymbol{r}}=m^{y}_{\boldsymbol{r}}=0,\qquad C^{x,z}_{\boldsymbol{r},\boldsymbol{r}^{\prime}}=C^{y,z}_{\boldsymbol{r},\boldsymbol{r}^{\prime}}=0.
\end{equation}
This does not hold for other initial conditions.

Upon closer inspection, the equal distance correlation functions
$C^{x,x}_{\boldsymbol{r}\boldsymbol{r}}=C^{y,y}_{\boldsymbol{r}\boldsymbol{r}}=1/4$ do not enter the cumulant equations. The set of coupled equations of motion can be written in compact form as
\begin{align}
    \label{eq:eom mz}
    \partial_{t}m^{z}_{\boldsymbol{r}} &=-2J(D C^{x,y})_{\boldsymbol{r}\boldsymbol{r}},
    \\
    \label{eq:eom cxx}
    \partial_{t} C^{x,x}_{\boldsymbol{r}\boldsymbol{r}^{\prime}}
    &=J m^{z}_{\boldsymbol{r}}
    \left(D C^{x,y}\right)_{\boldsymbol{r}\boldsymbol{r}^{\prime}}
    +J m^{z}_{\boldsymbol{r}^{\prime}}
    \left(D C^{x,y}\right)_{\boldsymbol{r}^{\prime}\boldsymbol{r}}
    \nonumber\\
    &\quad
    +\delta\!\left(t\right)
    \left(\eta_{\boldsymbol{r}}-\eta_{\boldsymbol{r}^{\prime}}\right)
    C^{x,y}_{\boldsymbol{r}\boldsymbol{r}^{\prime}},
    \\
    \label{eq:eom cxy}
    \partial_{t} C^{x,y}_{\boldsymbol{r}\boldsymbol{r}^{\prime}}
    &=-J m^{z}_{\boldsymbol{r}}
    \left(D C^{x,x}\right)_{\boldsymbol{r}\boldsymbol{r}^{\prime}}
    +J m^{z}_{\boldsymbol{r}^{\prime}}
    \left(D C^{x,x}\right)_{\boldsymbol{r}^{\prime}\boldsymbol{r}}
    \nonumber\\
    &\quad
    -\delta\!\left(t\right)
    \left(\eta_{\boldsymbol{r}}-\eta_{\boldsymbol{r}^{\prime}}\right) C^{x,x}_{\boldsymbol{r}\boldsymbol{r}^{\prime}}
    -\frac{J}{4}
    \left(m^{z}_{\boldsymbol{r}}-m^{z}_{\boldsymbol{r}^{\prime}}\right)
    D_{\boldsymbol{r}\boldsymbol{r}^{\prime}},
\end{align}
where $D_{\boldsymbol{r}\boldsymbol{r}^{\prime}}
=\left|\boldsymbol{r}-\boldsymbol{r}^{\prime}\right|^{-3}$ is the dipolar
kernel with vanishing diagonal and
$\delta\!\left(t\right)=\delta_{0}e^{-t/\tau}$ is the ramped staggered field.
The compact notation for the convolutions $\left(D C\right)_{\boldsymbol{r}\boldsymbol{r}^{\prime}}
=\sum_{\boldsymbol{s}}D_{\boldsymbol{r}\boldsymbol{s}} C_{\boldsymbol{s}\boldsymbol{r}^{\prime}}$ requires the convention of a vanishing diagonal, $C^{x,x}_{\boldsymbol{r}\boldsymbol{r}}
=C^{x,y}_{\boldsymbol{r}\boldsymbol{r}} \to 0$ (there is a cancellation for those constant terms).
Note that in case of periodic boundary conditions, there is an additional $\mathsf{T} R_x(\pi)$ symmetry, where $\mathsf{T}$ is a translational shift operator over a single site.
These equations are solved with a fourth-order Runge-Kutta method.


\section{Cumulant dynamics: results}
\label{sec:Dependence of Plateau}
\noindent

\begin{figure*}[t!]
  \centering

  \begin{minipage}[c]{0.04\textwidth}
    \centering
    \rotatebox[origin=c]{90}{OBC}
  \end{minipage}%
  \hspace{0.01\textwidth}%
  \begin{minipage}[c]{0.85\textwidth}
    \centering
    \makebox[\linewidth][l]{%
      \parbox[c]{\linewidth}{%
        \includegraphics[width=\linewidth]{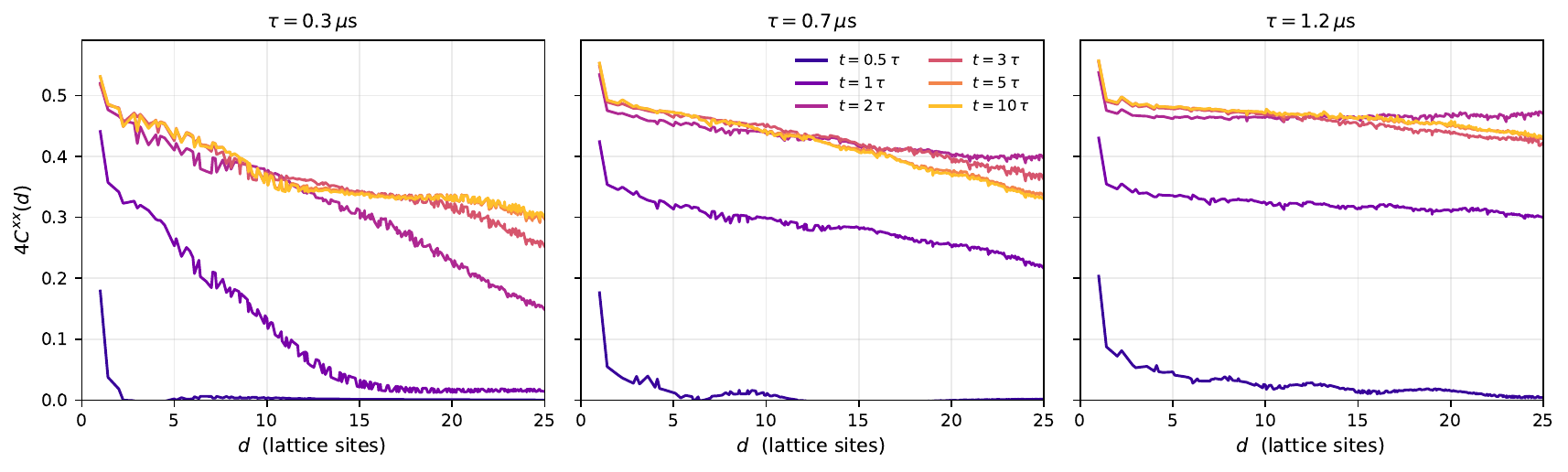}%
      }%
      \hspace{0.02\textwidth}%
      \parbox[c]{0.04\textwidth}{%
        \centering
        \rotatebox[origin=c]{90}{$s=0$}%
      }%
    }
    
    \vspace{0.25cm} 

    \makebox[\linewidth][l]{%
      \parbox[c]{\linewidth}{%
        \includegraphics[width=\linewidth]{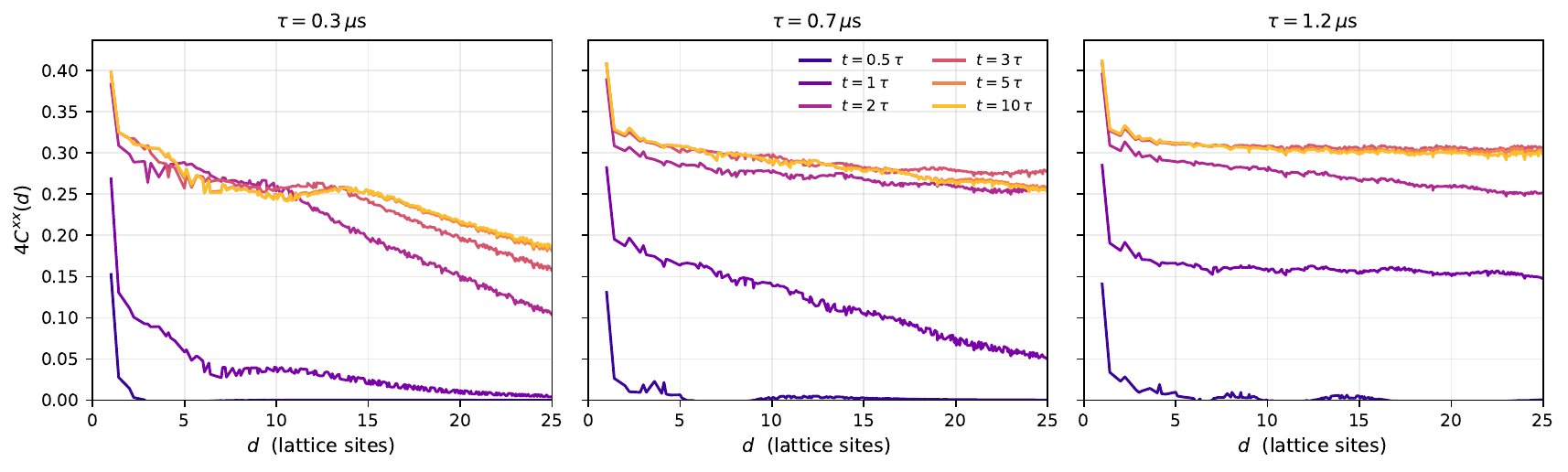}%
      }%
      \hspace{0.02\textwidth}%
      \parbox[c]{0.04\textwidth}{%
        \centering
        \rotatebox[origin=c]{90}{$s=0.3$}%
      }%
    }
  \end{minipage}

  \caption{\justifying
  Cumulant dynamics of the spatial spin correlation function $C^{x,x}$ for a system of linear size $L=50$ with open boundary conditions (OBC), prepared in the Néel state with zero initial entropy ($s=0$, top) or finite initial entropy ($s=0.3$, bottom).
  Each panel shows the evolution for a different ramp constant $\tau$.
  Note that the $x$-axis is truncated at half the linear system size (beyond which all correlation functions decay).
  }
  \label{fig:sub1}
  \label{fig:sub2}

  \vspace{0.25cm}

  \begin{minipage}[c]{0.04\textwidth}
    \centering
    \rotatebox[origin=c]{90}{PBC}
  \end{minipage}%
  \hspace{0.01\textwidth}%
  \begin{minipage}[c]{0.85\textwidth}
    \centering
    \makebox[\linewidth][l]{%
      \parbox[c]{\linewidth}{%
        \includegraphics[width=\linewidth]{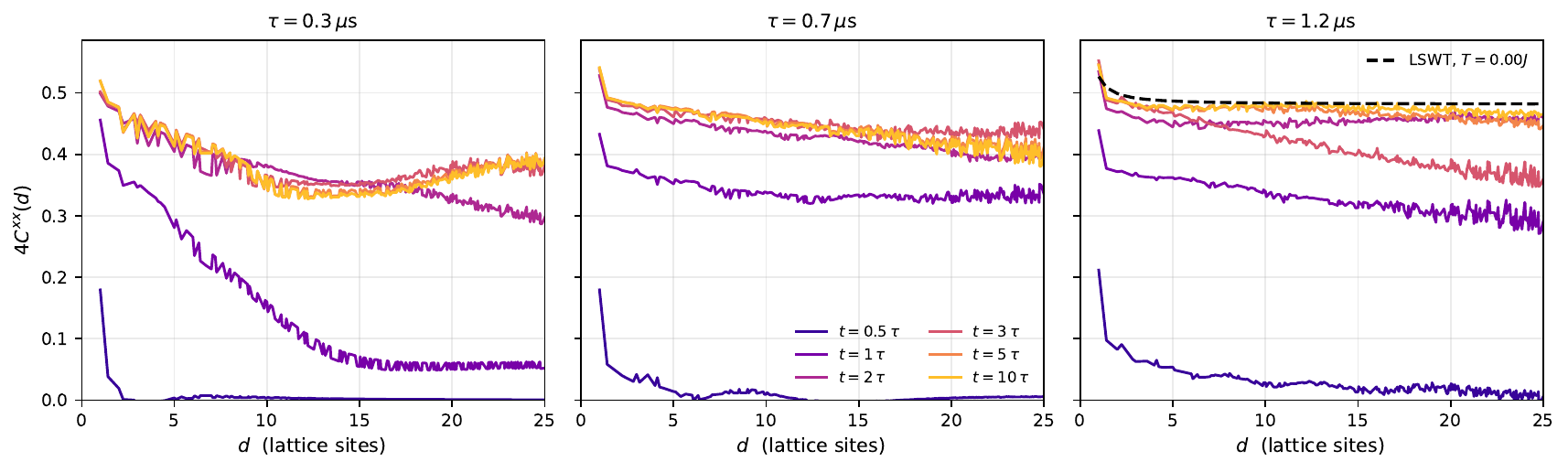}%
      }%
      \hspace{0.02\textwidth}%
      \parbox[c]{0.04\textwidth}{%
        \centering
        \rotatebox[origin=c]{90}{$s=0$}%
      }%
    }
    
    \vspace{0.25cm}

    \makebox[\linewidth][l]{%
      \parbox[c]{\linewidth}{%
        \includegraphics[width=\linewidth]{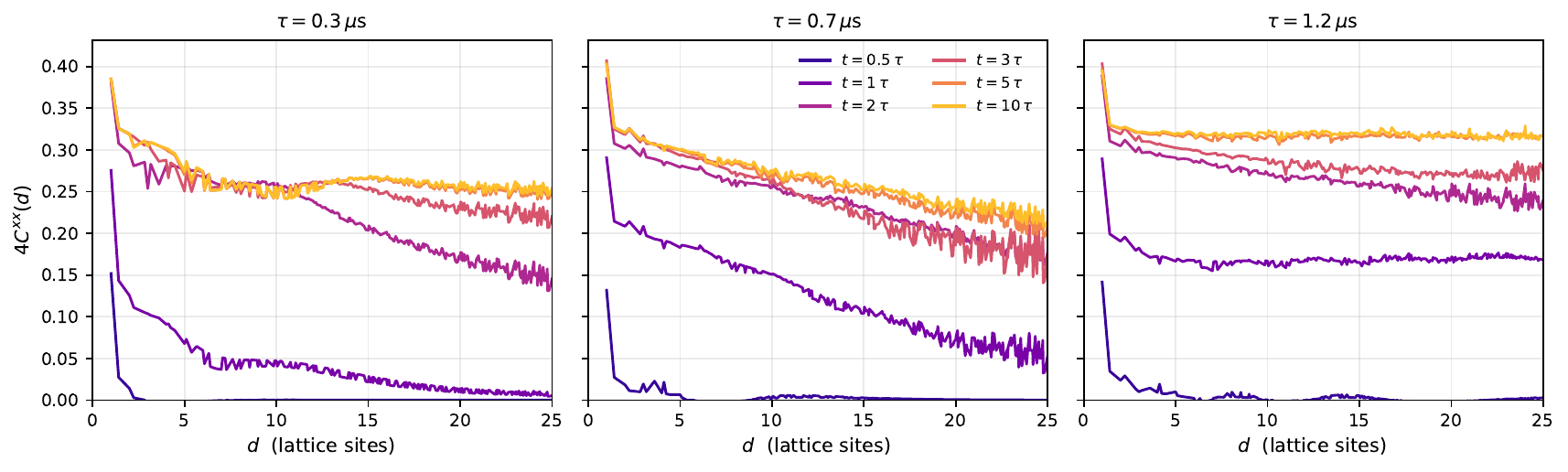}%
      }%
      \hspace{0.02\textwidth}%
      \parbox[c]{0.04\textwidth}{%
        \centering
        \rotatebox[origin=c]{90}{$s=0.3$}%
      }%
    }
  \end{minipage}

  \caption{\justifying
  Cumulant dynamics of the spatial spin correlation function $C^{x,x}$ for a system of linear size $L=50$ with periodic boundary conditions (PBC), prepared in the Néel state with zero initial entropy ($s=0$, top) or finite initial entropy ($s=0.3$, bottom).
  Each panel shows the evolution for a different ramp constant $\tau$.
  Note that the $x$-axis is truncated at half the linear system size because of mirror symmetry.
  The dashed curve in the right panel corresponds to the ground state LSWT prediction.
  }
  \label{fig:sub3}
  \label{fig:sub4}
\end{figure*}

Our main result is the cumulant dynamics for the spin correlation function $C^{x,x}$ as a function of boundary conditions (open: OBC; periodic: PBC), ramp decay time $\tau$, and entropy $s$ of the initial N\'eel state. Those are shown in~\Cref{fig:sub1,fig:sub2,fig:sub3,fig:sub4}, where we kept the strength $\delta_0$ fixed (which was found to be a less relevant parameter, primarily leading to shifts in time of the observed behavior) and worked at fixed linear system $L=50$ which is a system size that should become experimentally available in the foreseeable future. It also helps diminishing the effects of small system size. All times are reported in units of $t/\tau$.

The results can best be understood by first considering the role of the ramp constant $\tau$.

\subsection{The quasi-adiabatic regime:}
For $\tau$ sufficiently large -- see the right panels in~\Cref{fig:sub1,fig:sub2,fig:sub3,fig:sub4} --  the boundary conditions do not matter much, and a quasi-plateau is formed in all cases for sufficiently long times. Every $\tau$ supports a certain maximum extension of the plateau. The plateau also sets in at very short distances, $d\sim2$. For the ground state, it is very close to the predictions of LSWT.  Entropy is seen to reduce the value of the plateau (\textit{i.e.,} the spontaneously formed in-plane magnetization) but we can no longer compare with LSWT because an initial state with $s=0.3$ leads to temperatures outside the validity of LSWT~\cite{Sbierski2024Magnetism}.

\subsection{The diabatic regime:}
When $\tau$ is short -- see the left panels in~\Cref{fig:sub1,fig:sub2,fig:sub3,fig:sub4} -- the system is far away from equilibrium. Boundary conditions play a role but a minor one, even for $L=50$, especially at the later times: The $s=0$ spin correlation function oscillates stronger for PBC than for OBC, whereas for $s=0.3$ the OBC spin correlation function decays much faster for $d>15$ than the PBC one. The most interesting feature is the formation at moderate times of a shoulder at intermediate distances, $10 < d < 20$ for $s=0$ and  $5 < d < 15$ for $s = 0.3$, which, in the PBC case, has an additional small modulation. 
The shoulder is the diabatic system's best possible effort in mimicking a $k=0$ zero mode.
The onset value of this shoulder is very different than the one of an equilibrium plateau, showing stronger short-range relaxation effects. The shoulder survives to long times. Entropy leads to quantitative changes (greater onset time, lower value of the plateau for $s=0.3$ vs.~$s=0$). The extent of the shoulder does not scale with $L$ for large enough~$L$.

\begin{figure*}[t!]
  \centering
  \includegraphics[width=\textwidth]{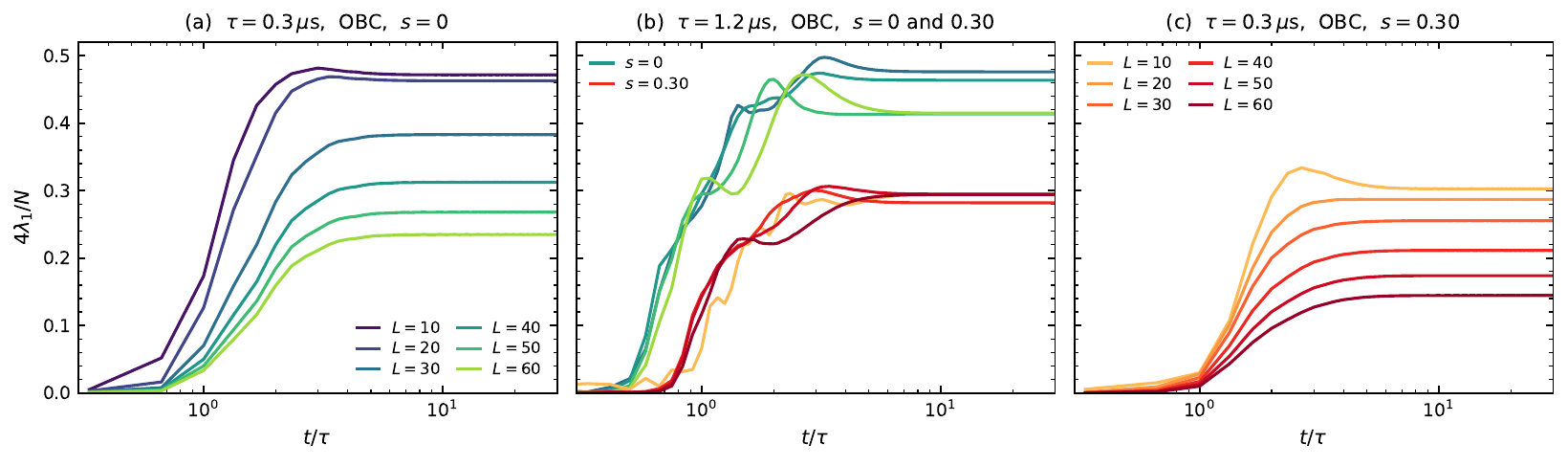}
  \caption{\justifying
  Largest eigenvalue of the $C^{x,x}$ correlation function as a function of time for different ramp constants, initial entropies, and system sizes with open boundary conditions (OBC). Left panel: fast ramp in the ground state, showing the absence of ODLRO. Middle: slow ramp in the ground state and at finite entropy, showing genuine ODLRO. Right: fast ramp at finite entropy, showing the absence of ODLRO.}
  \label{fig:lambda1_time_Neel}
\end{figure*}

\begin{figure*}
\includegraphics[width=\textwidth]{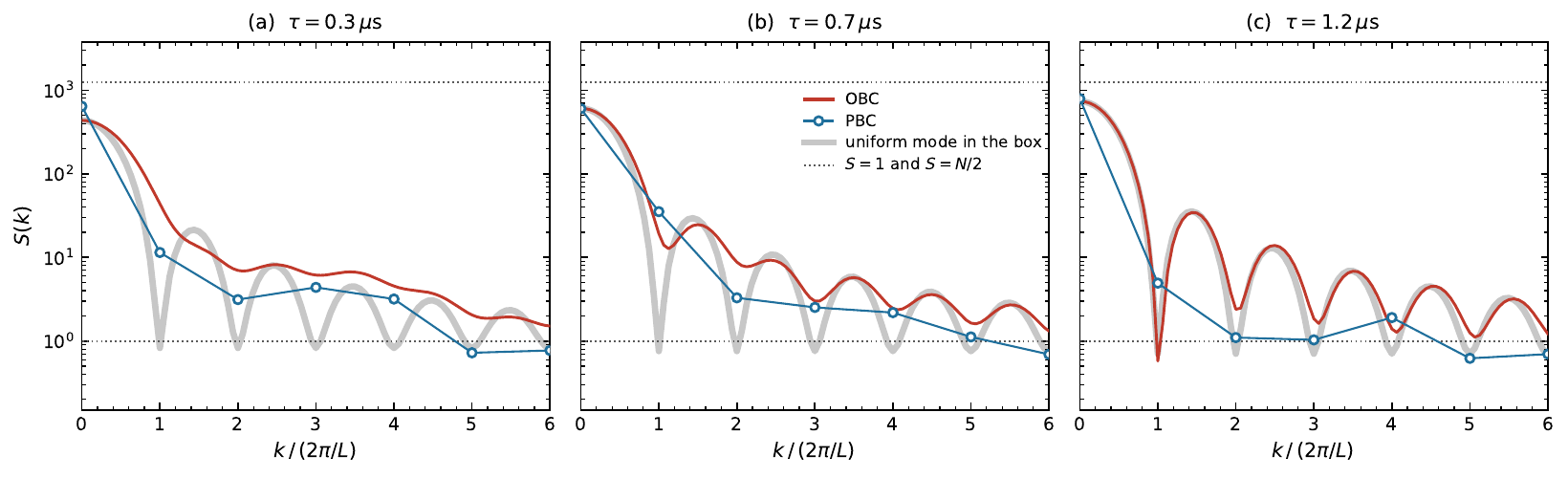}
\caption{\label{fig:sk}%
\justifying
Transverse structure factor
$S(\mathbf{k})=N^{-1}\sum_{ij}e^{i\mathbf{k}\cdot(\mathbf{r}_i-\mathbf{r}_j)}
 \langle C^{x,x}_{\mathbf{r}_i, \mathbf{r}_j} \rangle$
along the axial cut $\mathbf{k}=(k,0)$ at $L=50$, $s=0.30$ and $t=10\tau$, for
(a)~$\tau=0.3\,\mu$s, (b)~$0.7\,\mu$s and (c)~$1.2\,\mu$s.  The dotted lines $S=1$ and $S(0)=N/2$ correspond to the uncorrelated and the $T=0$ product-state ferromagnet $4C^{xx}=1/2$, respectively. Red lines are for OBC, blue lines are a guide to the eye for PBC. The grey curve is the zero mode for OBC, adjusted at $k=0$ (see text). 
}
\end{figure*}

We also checked the scaling of the largest eigenvalues of the spin correlation functions, and found that the largest one scales sub-extensively $\sim L^{1.4}$ as shown in Fig.~\ref{fig:lambda1_time_Neel}.
Per the Penrose criterion~\cite{Penrose1956,Yang1962}, this rules out ODLRO. The eigenvector corresponding to the largest eigenvalue is the $k=0$ mode for PBC. For OBC it also has a strong overlap ($\gtrsim 99\% $ for $L=30$) with it, though this fraction goes down with system size. For OBC, the subleading eigenvalues of $C^{x,x}$ grow super-extensively; they are responsible for the shoulder termination. It is indeed in the low-$k$ structure how PBC and OBC differ, as can be seen in Fig.~\ref{fig:sk}.
For PBC (blue dots), momentum is a good quantum number and $S(\mathbf{k})$ is the eigenvalue spectrum of $4C^{xx}+1$ (the blue line is a guide to the eye, only the dots have meaning). The grey curve shows a perfectly condensed
uniform mode (zero mode) for OBC
\begin{equation}
S_{\rm box}(\mathbf{k})=1+4m^2(|F(\mathbf{k})|^2-N)/N,
\label{eq:zero_mode_structure}
\end{equation}
with
\begin{equation}
F(\mathbf{k})=\sum_ie^{i\mathbf{k}\cdot\mathbf{r}_i}.
\label{eq:F}
\end{equation}
Its scale is fixed through the $k=0$ value of the red curve by setting $4m^2=(S(0)-1)/(N-1)$.
The ramps differ by the minima in the structure factor; OBC differs from PBC by filling in $k$-values between the minima.  Slow ramps reproduce the OBC zero mode structure to within $20\%$, while the experimental ramp exceeds it by a factor $\sim 50$. The shoulder termination for $\tau=0.3\,\mu$s takes place at $k/(2\pi/L) \approx 1.5$ in Fig.~\ref{fig:sk}a, which is at a maximum for the OBC curve and absent for the PBC curve (see also Sec.~\ref{sec:Cross-Validation and Positivity}).

\subsection{The intermediate regime:}
For intermediate ramp constants -- see the middle panels in ~\Cref{fig:sub1,fig:sub2,fig:sub3,fig:sub4} -- shoulder formation disappears. For OBC the system is close to forming a plateau after substantial short-distance relaxation, but is unable to sustain the plateau at larger times. PBC shows a similar overshoot at $s=0.3$, whereas $s=0$ case seems close to a genuine plateau only 20$\%$ below the thermodynamic one. 

\subsection{Implications for the experimental observations}
The observed behavior at $s=0.3$ is consistent with experimental findings~\cite{Chen2023ContinuousSymmetryBreaking,Chen2025QuenchSpectroscopy}: Experimentalists found consistently this flattening for different repetitions of the experiment and initial configurations on $L=6$ and $L=10$ (OBC). The results of~\Cref{fig:sub1,fig:sub2,fig:sub3,fig:sub4} show that the flattening must be interpreted as a non-extensive shoulder and Fig.~\ref{fig:sub2} (bottom) indicates that the lack of adiabaticity in experiment is the main reason for shoulder formation under coherent dynamics. We note that in experiment the ramp constant was chosen as a tradeoff between adiabaticity and lifetime concerns. Experiment yielded indistinguishable results for $\tau = 0.15\,\mu$s and $\tau=0.3\,\mu$s.  It was argued that the loss of adiabaticity is not dominant versus other imperfections such as black-body radiation, spontaneous emission, addressing laser issues, jitter. Here we see that the required time constant for studying adiabatic physics was underestimated.  The too small experimental system size masks the loss of adiabaticity.  The imperfections in the system lead to thermalization of the experiment at much later times, cf.~the arguments presented in Ref.~\cite{Sbierski2024Magnetism}.


\section{Cross-Validation and Positivity}
\label{sec:Cross-Validation and Positivity}
\noindent

Although the cumulant dynamics seems to capture many aspects of the experiment in Ref.~\cite{Chen2023ContinuousSymmetryBreaking}, the dynamics is not exact because it neglects irreducible three-body correlations, which breaks the exact spin algebra and must therefore lead to unphysical behavior at long times. In this section we examine these artifacts more closely.

The internal physical consistency of the formalism can be checked by relations such as
\begin{eqnarray}
\label{eq:dynamical consistency conditions of ramped dipolar XY model}
C^{a,b}_{\boldsymbol{r},\boldsymbol{r}^{\prime}}=C^{b,a}_{\boldsymbol{r}^{\prime},\boldsymbol{r}} & : & {\rm hermiticity} \nonumber \\
\left\langle\hat{S}^{z}_{\text{total}}\right\rangle=\text{constant}\quad\left(\delta=0\right) & : & {\rm unitarity} \nonumber \\
\vert m^{x,y,z} \vert \le 1/2 & : & {\rm embedding} \nonumber \\ 
\lambda_{\min}\left(\rho_{\boldsymbol{r},\boldsymbol{r}^{\prime}}\right)\geq0 & : & {\rm embedding} \nonumber \\
\end{eqnarray}
Here, $\lambda$ denotes eigenvalues.
Hermiticity is explicitly respected in our formalism, as is the conservation of total magnetization and energy. 

The two embedding conditions are related to the fact that the formalism is not working directly on the space of physical density matrices but rather on cumulants from which the density matrices need to be reconstructed.
The two-body density matrix, see Eq.~\eqref{eq:definition of two-body density function of ramped dipolar XY model}, should have the properties, 
\begin{equation}
\label{eq:requirement for two-body density matrix of ramped dipolar XY model}
\rho^{\dagger}_{\boldsymbol{r},\boldsymbol{r}^{\prime}}=\rho_{\boldsymbol{r},\boldsymbol{r}^{\prime}},\qquad{\rm Tr}(\rho_{\boldsymbol{r},\boldsymbol{r}^{\prime}}) = 1,\qquad\lambda\left[\rho_{\boldsymbol{r},\boldsymbol{r}^{\prime}}\right]\geq0.
\end{equation}
Hermiticity is obvious. The trace condition follows immediately from the zero trace of the spin operators. But the semi-positive definiteness of the two-body density matrix will not necessarily be satisfied at all times. We denote  $\lambda_{\min}\left(t\right)=\min_{\boldsymbol{r}\neq\boldsymbol{r}^{\prime}} \lambda\left[\rho_{\boldsymbol{r},\boldsymbol{r}^{\prime}}\right]$ its minimum eigenvalue. For the ramps discussed in the previous section initialized, we find that $\lambda_{\min}$ goes occasionally negative ($\lesssim -0.01$) after long times but only when the initial entropy is close to zero. For initial states with higher entropy, positivity was respected for all times.  
The local magnetization also stays within bounds at all times and sites. 
We have compared the cumulant dynamics on a system of linear size $L=4$ with open boundary conditions with the exact ones obtained by full diagonalization (as well as with the time-dependent variational Monte Carlo simulation of Ref.~\cite{Chen2025QuenchSpectroscopy} for a system of linear size $L=10$). The exact dynamics are only followed up to a time $t \sim 0.7\,\mu$s, after which a pointwise comparison shows large relative deviations. Those are the largest for short distances and grow with time. As we noted, quantitative comparisons are beyond reach for this formalism, but the experiment's qualitative features seem to be captured.

This brings us to the pertinent question why long-time cumulant dynamics of plateaus and shoulders could be physical. 

In the quasi-adiabatic regime, Fig.~\ref{fig:sub3} shows good agreement with LSWT at long times for $s=0$. The reason is that the true ordered ground state is very close to a product state~\cite{Block_2024,KaplanLipkin2025}. The depletion, which controls the neglected three-body correlations is very small. The near-agreement with LSWT is thus kinematically enforced. The case $s=0.3$ is still in the deeply ordered phase, even though LSWT no longer applies, and the same kinematic constraint applies. 

The diabatic regime, $\tau = 0.3$, requires a different argument. The short-range physics is strongly entangled and beyond control of the cumulant dynamics. But all observations at intermediate and long distances can be captured by the following model:
\begin{equation}
C^{xx}_{\rm model}(r,t)  = \frac{m^{2}}{1+\left(r/r_{c}\right)^{3}}.
\label{eq:model}
\end{equation}
The amplitudes $m(\tau,s,t)$ and the crossover length
$r_{c}(\tau,s,t)$ are both functions of the ramp time, the initial entropy and the observation time, and are provided by the cumulant dynamics. The $\sim a/r^3$ paramagnetic decay at large distances with $a = m^{2}r_{c}^{3}$ is set by the dipolar interactions. As $\lambda_1 \sim m^2 r_c^2$, there will be no ODLRO unless $r_c \sim L$ (which is the adiabatic regime when $L$ is very large). 
The convergence to $\lambda_1$ in the thermodynamic limit is slow because of the dipolar interactions, and will be a factor $\ln(L)$ slower for open than for periodic boundary conditions.

The implication of the model is that for ramps like $\tau = 0.3\,\mu$s that are neither too fast nor too slow, a coherent region with a finite magnetization can be supported up to some length scale. The correlations remain paramagnetic (and thus dipolar) outside this region. Such a region can be quite large, though does not scale with system size, and its magnetization can also be surprisingly large. The cumulant dynamics show that this state is very long-lived, though it certainly does not correspond to the thermal state, and is also not physical: The failure of thermalization under the system's own dynamics is an artifact of the cumulant dynamics and can be understood as follows: when $m_z \to 0$ and $\delta \to 0$, then effectively $\partial_t C \to 0$. The cumulant dynamics neglects for instance magnon-magnon scattering, which contributes to the neglected third order spin correlations. Those terms remain small, and it is therefore entirely plausible that the experiment has observed such a shoulder that extends over almost the full system after $t\sim 1\,\mu$s. Crudely estimating such a decay at $T=0.5J$ for an equilibrium density of spin waves described by LSWT yields relaxation coefficients $\Gamma_k / J \ll \mathcal{O}(10^{-2} \, - \, 10^{-3})$ at $k = 2\pi/L$ (and smaller $k$).

\section{Spectroscopy through dynamical preparation}
\label{sec:prep_y}

Ref.~\cite{Chen2025QuenchSpectroscopy} aimed at performing spectroscopy through dynamical state preparation.
The initial state was however no longer a N\'eel state in an external staggered magnetic field, but it consisted of all spins oriented in the $\hat{y}-$direction,
\begin{equation}
\label{eq:initial value of magnetization in cross validation of dipolar XY model}
m^{y}_{\boldsymbol{r}}\left(0\right)=\frac{1}{2}, \quad  m^{x}_{\boldsymbol{r}}\left(0\right)=m^{z}_{\boldsymbol{r}}\left(0\right)=0, \quad  C^{a,b}_{\boldsymbol{r} \neq \boldsymbol{r}^{\prime}}\left(0\right)=0.
\end{equation}
As before, onsite quantum fluctuation do not vanish.

The cumulant dynamics for the initial N\'eel state made use of the $U(1)$-symmetry around the $z$-axis. A state prepared in the $xy$ plane breaks that symmetry. In the absence of a staggered field, a $\pi$ rotation about $y$,
$\left(S^{x},S^{y},S^{z}\right)\rightarrow\left(-S^{x},S^{y},-S^{z}\right)$,
leaves however the interaction invariant. It gives
$m^{x}=m^{z}=0$ and $C^{x,y}=C^{y,z}=0$, while
$C^{x,z}$ survives. The closed set of equations is correspondingly larger:
\begin{align}
\label{eq:eom my}
\partial_{t}m^{y}_{\boldsymbol{r}}
&=J\sum_{\boldsymbol{s}}D_{\boldsymbol{r}\boldsymbol{s}}\,
C^{x,z}_{\boldsymbol{s}\boldsymbol{r}},
\\
\label{eq:eom cxx y}
\partial_{t}C^{x,x}_{\boldsymbol{r}\boldsymbol{r}^{\prime}}
&=-J\Lambda^{\boldsymbol{r}^{\prime}}_{\boldsymbol{r}}\,
C^{x,z}_{\boldsymbol{r}^{\prime}\boldsymbol{r}}
-J\Lambda^{\boldsymbol{r}}_{\boldsymbol{r}^{\prime}}\,
C^{x,z}_{\boldsymbol{r}\boldsymbol{r}^{\prime}},
\\
\label{eq:eom cxz}
\partial_{t} C^{x,z}_{\boldsymbol{r}\boldsymbol{r}^{\prime}}
&=-\frac{J}{4}D_{\boldsymbol{r}\boldsymbol{r}^{\prime}}m^{y}_{\boldsymbol{r}^{
\prime}}
-J\Lambda^{\boldsymbol{r}^{\prime}}_{\boldsymbol{r}}\,
C^{z,z}_{\boldsymbol{r}\boldsymbol{r}^{\prime}}
\nonumber\\
&\quad
+J\!\!\sum_{\boldsymbol{s}\neq\boldsymbol{r},\boldsymbol{r}^{\prime}}\!\!
D_{\boldsymbol{r}^{\prime}\boldsymbol{s}}
\Bigl(m^{y}_{\boldsymbol{s}}
C^{x,x}_{\boldsymbol{r}\boldsymbol{r}^{\prime}}
-m^{y}_{\boldsymbol{r}^{\prime}}
C^{x,x}_{\boldsymbol{s}\boldsymbol{r}}\Bigr),
\\
\label{eq:eom czz}
\partial_{t}C^{z,z}_{\boldsymbol{r}\boldsymbol{r}^{\prime}}
&=J\!\!\sum_{\boldsymbol{s}\neq\boldsymbol{r},\boldsymbol{r}^{\prime}}\!\!
D_{\boldsymbol{r}\boldsymbol{s}}
\Bigl(m^{y}_{\boldsymbol{s}}
C^{x,z}_{\boldsymbol{r}\boldsymbol{r}^{\prime}}
-m^{y}_{\boldsymbol{r}}
C^{x,z}_{\boldsymbol{s}\boldsymbol{r}^{\prime}}\Bigr)
\nonumber\\
&\quad
+J\!\!\sum_{\boldsymbol{s}\neq\boldsymbol{r},\boldsymbol{r}^{\prime}}\!\!
D_{\boldsymbol{r}^{\prime}\boldsymbol{s}}
\Bigl(m^{y}_{\boldsymbol{s}}
C^{x,z}_{\boldsymbol{r}^{\prime}\boldsymbol{r}}
-m^{y}_{\boldsymbol{r}^{\prime}}
C^{x,z}_{\boldsymbol{s}\boldsymbol{r}}\Bigr),
\end{align}
where
\begin{equation}
\label{eq:lambda def}
\Lambda^{\boldsymbol{r}^{\prime}}_{\boldsymbol{r}}
\equiv\!\!\sum_{\boldsymbol{s}\neq\boldsymbol{r},\boldsymbol{r}^{\prime}}\!\!
D_{\boldsymbol{r}\boldsymbol{s}}m^{y}_{\boldsymbol{s}}
=\left(Dm^{y}\right)_{\boldsymbol{r}}
-D_{\boldsymbol{r}\boldsymbol{r}^{\prime}}m^{y}_{\boldsymbol{r}^{\prime}}.
\end{equation}
We stress again that the equations only hold when the $\pi$ rotation about $y$ is a symmetry.
Note that the transverse and longitudinal correlations evolve together: The second term of Eq.~\eqref{eq:eom cxz} feeds $C^{z,z}$ into $C^{x,z}$ whereas in the N\'eel sector $C^{x,z} \equiv 0$.

The transverse correlator $C^{y,y}$ takes the role that
$C^{z,z}$ played in the N\'eel sector: it is generated but does not feed back. It is only needed if it is measured $ \partial_{t}C^{y,y}_{\boldsymbol{r}\boldsymbol{r}^{\prime}}
=-JD_{\boldsymbol{r}\boldsymbol{r}^{\prime}}
\Bigl(m^{y}_{\boldsymbol{r}^{\prime}}
C^{x,z}_{\boldsymbol{r}^{\prime}\boldsymbol{r}}
+m^{y}_{\boldsymbol{r}}
C^{x,z}_{\boldsymbol{r}\boldsymbol{r}^{\prime}}\Bigr)
$, which can be evaluated cheaply. It must however be included in the energy calculation. At short times, $C^{x,z}_{\boldsymbol{r},\boldsymbol{r}^{\prime}}$ is linear in time, while $C^{x,x}_{\boldsymbol{r},\boldsymbol{r}^{\prime}}$ and $C^{z,z}_{\boldsymbol{r},\boldsymbol{r}^{\prime}}$ are quadratic in time.
The full dynamics remains strictly hermitian, conserves the total and local $z-$magnetization (trivially) and energy at all times. The density matrix suffers from the same negative eigenvalue problem as reported for the N\'eel state preparation. We also found that $\vert m^y\vert$ can occasionally exceed $1/2$ by a tiny amount, but only when the initial state is fully polarized along $\hat y$.

\begin{figure}[!ht]
    \includegraphics[width=\columnwidth]{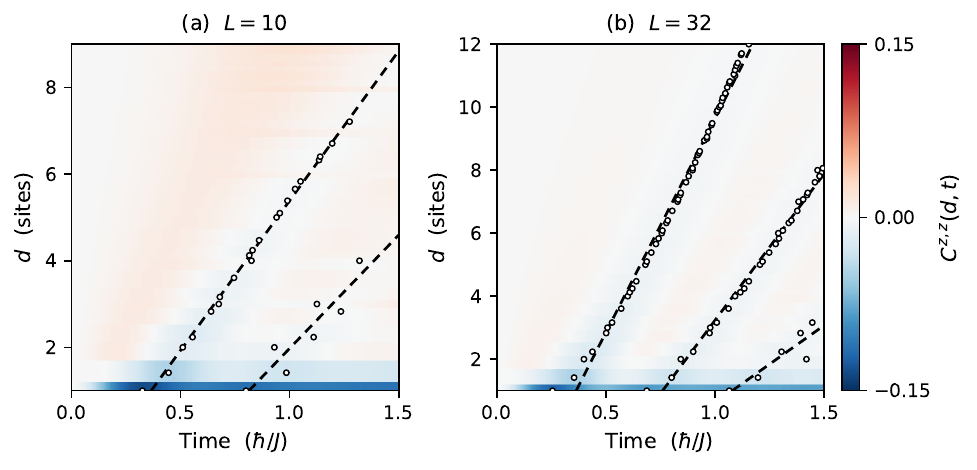}
    \caption{\justifying
    The longitudinal correlation
    $C^{z,z}_{\sigma}(d,t)$ as a function of space and time following the protocol of Ref.~\cite{Chen2025QuenchSpectroscopy} with open boundaries  for (a)~$L=10$, and (b)~$L=32$. Time is in the same units as the experiments (factor 2 difference in  $\hbar/J$). 
    Dashed lines are ballistic fits to the local minima (circles) of $C^{z,z}_{\sigma}$.}
    \label{fig:sprectral_czz}
\end{figure}

In Fig.~\ref{fig:sprectral_czz} we show the longitudinal correlation function $C^{z,z}_{\sigma}$ as a function of space time following where the initial state is prepared along $\hat{y}$. The cumulant dynamics show the propagation of several wavefronts, corresponding to the generation of magnon pairs in the quench dynamics. Ref.~\cite{Chen2025QuenchSpectroscopy} showed the corresponding experimental data, which are very close to the ones shown in the left panel of Fig.~\ref{fig:sprectral_czz}. They overlaid the data with a $d\sim t^{2}$ curve, which follows from the the dispersion $\omega\sim\sqrt{k}$ in the thermodynamic limit. The longest wavelengths in our finite system with open boundary conditions are  $k_{\min}=\pi\sqrt{2}/(L+1)$ and those determine the group velocity of the first wavefront which is hence ballistic.

\section{Conclusions}
\label{sec:Conclusions}

We have analyzed the dynamics of Refs.~\cite{Chen2023ContinuousSymmetryBreaking,Chen2025QuenchSpectroscopy} using cumulant expansions truncated to second order. The method is not exact for long times, but it does capture the intermediate time scales reasonable well and can explain the experimental puzzles. In the staggered magnetic field setup, the ramp constant is the most important quantity: for the ramps used in the experiment, $\tau = 0.3\,\mu$s, the system can support a region with a finite (but weak) magnetization up to surprisingly long distances; in fact, longer than the currently available system sizes, and this explains the experimentally observed plateau in $C^{x,x}$ at times $t \sim 1-2\,\mu$s on $L=10$. It is however not a signal of ODLRO because for bigger system sizes the plateau in the correlation function develops into a shoulder followed by a dipolar decay at larger distances. Within cumulant dynamics, this shape is the one of the steady-state though in reality the omitted higher order processes will lead to damping and a {\it very} slow thermalization of the system. In order to reach a plateau extending over $L \sim 50$ sites, the model predicts that a ramp constant of at least $\tau \sim 1.2\,\mu$s will be needed, and that this in turn will require longer experimental lifetimes. The cumulant dynamics could likewise reproduce the quench spectroscopy of Ref.~\cite{Chen2025QuenchSpectroscopy} well. The role played by the open boundary conditions remains remarkably strong, even at $L=32$. Near-future experiments in which the system length and time constant are systematically scanned over can falsify our proposed scenario.

\begin{acknowledgments}

Large-language models were used to generate the figures, optimize the Python code for the cumulant dynamics, obtain order of magnitude estimates, and check/improve the text. We used to this end Claude Opus 5, OpenAI ChatGPT 5.5, and Grammarly.

We wish to thank A. Browaeys, L. Homeier, B. Sbierski, and N. Yao for stimulating discussions.
We acknowledge financial support from the Deutsche Forschungsgemeinschaft (DFG), project Nr 530111096.
\end{acknowledgments}

\bibliography{literature}
\end{document}